\documentclass[12pt,twoside]{article}  
\usepackage{natbib}
\usepackage{fancyhdr}		

\usepackage[section]{placeins}   %

\usepackage{graphicx}

\newcommand{\cen}[1]{\begin{center} #1 \end{center}}

\usepackage[]{lineno}

\usepackage{setspace}
\usepackage{amssymb}
\usepackage{graphicx}
\usepackage{subfigure}
\usepackage{amsmath}
\usepackage{multirow}
\usepackage{booktabs}
\usepackage{dcolumn}
\usepackage{enumerate}
\usepackage{authblk}
\usepackage{makecell}
\usepackage[font=small,labelfont=bf]{caption}
\usepackage[colorlinks=true,linkcolor=blue,urlcolor=black,bookmarksopen=true,citecolor=red,bookmarks=true]{hyperref}

\begin{document}

\cen{\sf {\Large {\bfseries An Analysis of Scatter Characteristics in X-ray CT Spectral Correction} \\  
\vspace*{10mm}
Tao Zhang$^{1,2}$, Zhiqiang~Chen$^{1,2}$, Hao~Zhou$^{1,2}$, N.~Robert~Bennett$^{3}$, Adam~S.~Wang$^{3}$, and Hewei~Gao$^{1,2*}$} \\
\vspace*{5mm}
$^{1}$Department of Engineering Physics, Tsinghua University, Beijing 100084, China\\
$^{2}$Key Laboratory of Particle \& Radiation Imaging (Tsinghua University), Ministry of Education, China\\
$^3${Department of Radiology, Stanford University, Stanford, CA 94305, USA}\\
}

\setcounter{page}{1}
\pagestyle{plain}
*Author to whom correspondence should be addressed. Email: hwgao@tsinghua.edu.cn \\
\begin{abstract}
\indent
X-ray scatter remains a major physics challenge in volumetric computed tomography (CT), whose physical and statistical behaviors have been commonly leveraged in order to eliminate its impact on CT image quality. 
In this work, we conduct an in-depth derivation of how the scatter distribution and scatter to primary ratio (SPR) will change during the spectral correction, leading to an interesting finding on the property of scatter: when applying the spectral correction before scatter is removed, the impact of SPR on a CT projection will be scaled by the first derivative of the mapping function; while the scatter distribution in the transmission domain will be scaled by the product of the first derivative of the mapping function and a natural exponential of the projection difference before and after the mapping.
Such a characterization of scatter's behavior provides an analytic approach of compensating for the SPR as well as approximating the change of scatter distribution after spectral correction, even though both of them might be significantly distorted as the linearization mapping function in spectral correction could vary a lot from one detector pixel to another.
We conduct an evaluation of SPR compensations on a Catphan phantom and an anthropomorphic chest phantom to validate the characteristics of scatter. 
In addition, this scatter property is also directly adopted into CT imaging using a spectral modulator with flying focal spot technology (SMFFS) as an example to demonstrate its potential in practical applications.
For cone-beam CT scans at both 80 and 120 kVp, CT images with accurate CT numbers can be achieved after spectral correction followed by the appropriate SPR compensation based on our presented scatter property.
In the case of the SMFFS based cone-beam CT scan of the Catphan phantom at 120 kVp, after a scatter correction using an analytic algorithm derived from the scatter property, CT image quality was significantly improved, with the averaged root mean square error reduced from 297.9 to 6.5 HU.
\vspace{5mm}
\end{abstract}



\newpage
\setlength{\baselineskip}{0.7cm}      
\pagestyle{fancy}
\section{Introduction}\label{sec:intro}
X-ray scatter poses a great challenge in volumetric computed tomography (CT) such as wide-coverage multi-detector CT (MDCT) and flat panel-detector based cone-beam CT
(CBCT)\citep{hsieh2015computed,siewerdsen2001cone}, since the magnitude of scatter increases as the irradiation volume increases and the scatter signals can easily cause shading, cupping, and streak artifacts, as well as CT number inaccuracy in reconstructed images\citep{joseph1982effects}. As a result, compensation for or estimation of scatter is increasingly important in state-of-the-art CT imaging\citep{Ruhrnschopf2011}, in which the physical and statistical behaviors of scatter are commonly leveraged in scatter correction.

In medical applications, physical behaviors of scatter in CT can be characterized from how an X-ray photon interacts with matter in the effect of Compton scattering, whose differential cross section fits a model established by the Klein–Nishina formula\citep{klein1929streuung}. The Klein-Nishina formula may be directly applied to scatter compensation in single photon emission computed tomography\citep{bai2000slice}. In X-ray CT, however, one can take advantage of the facts that scattered X-ray photons carry lower energies and that they travel in directions away from that of the primary ones, to either distinguish them by their energy levels or reject them by a well designed anti-scatter grid\citep{Melnyk2014}, beam stop array\citep{Ning2004_BSA}, or even air-gap\citep{Sorenson1985}. Meanwhile, software tools based on Monte Carlo methods have been successfully developed to simulate the physical behavior of X-ray photons when scanning an object and estimate the corresponding scatter profiles\citep{xu2015practical}. Such an estimate can be extremely accurate as long as statistical convergence is met. To achieve real-time scatter correction, a deep convolution neural network has been trained to predict Monte Carlo scatter based on the acquired projection data \citep{maier2019real}. Recent studies have demonstrated that solving the Boltzmann transport equation can be an alternative way of accurately computing the distribution of scatter with much better computational performance\citep{Maslowski2018_Boltzmann,Wang2018_Boltzmann}. 
On the other hand, it is well-known that low-frequency components are usually dominant in the scatter distribution, which also has a relatively weak correlation with its primary. These statistical behaviors have played critical roles in kernel based scatter estimation algorithms \citep{sun2010improved,nomura2020modified,kim2015data}, the modulator based scatter correction method \citep{zhu2006scatter,gao2010scatter,bier2017scatter,pivot2020scatter}, and many others, directly or indirectly. Due to the inherent randomness in the production of X-ray photons, low-frequency components of the scatter distribution may be well corrected but not its noise. As a result, after scatter correction it is common to observe some increase in noise, which can be further suppressed by an appropriate noise reduction scheme\citep{Zhun2009_NoiseReduction}. In addition, scatter-to-primary ratio (SPR) is an important factor to assess the impact of scatter on the reconstructed CT image. However, unlike the scatter distribution, the SPR no longer can be considered by default as dominated by low-frequency components.
Joseph and Spital found that, for large body parts, scatter effects dominate over spectrum beam hardening effects primarily because of the increase in the SPR \citep{joseph1982effects}. It is also known that SPR increases as the field of view expands in the cone angle direction \citep{siewerdsen2001cone}. Also, an investigation on a dedicated cone-beam beast CT indicated that the measured SPR can be slightly affected by X-ray beam spectrum and object compositions \citep{kwan2005evaluation}. As a matter of fact, hardening of the X-ray beam spectrum (namely beam hardening) is another major source of error causing artifacts and inaccuracy in CT imaging \citep{brooks1976beam}, and a spectral correction (including but not limited to correction for beam hardening of the scanned object) is needed before final image reconstruction\citep{hsieh2015computed}

In general, scatter correction is done in the transmission domain before the negative logarithm operation, and the spectral correction is usually carried out after the removal of scatter which typically is a linearization mapping in the projection domain for a polychromatic CT system.
However, applying beam hardening correction before scatter is removed is desired as well in some cases.
One good example is in the scatter correction method for the primary modulator, where beam hardening of the modulator significantly affects the performance of scatter estimation \citep{gao2010modulator}. An inner-loop correction approach, where scatter and beam hardening are modeled and corrected within a unified framework of iterative image reconstruction, can be another example \citep{nuyts2013modelling}. 

In this work, we characterize a new scatter property that can play an important role in scenarios where a scatter correction is coupled with a spectral correction. As we know in medical CT applications, it is observed that most SPR is less than 200\%, and also that the linearization mapping function in the CT spectral correction for a water-equivalent object can be roughly approximated by a low-order Taylor expansion. Using the above two physical phenomena in practice, we conduct an in-depth analysis of how the scatter distribution and its SPR will change during the spectral correction, and hence obtain a new property of scatter: \textit{when applying the spectral correction before scatter is removed, the impact of SPR on the CT projection will be scaled, roughly, by the first derivative of the mapping function; while the scatter distribution in the transmission domain will be scaled, roughly, by the product of the first derivative of the mapping function and a natural exponential of the projection difference before and after the mapping}. Such a characterization of scatter property provides an analytic approach of compensating for the SPR as well as approximating the change of scatter distribution after spectral correction, even though both of them might be significantly distorted since the linearization mapping function may vary a lot from one detector pixel to another. To validate our proposed characteristics of scatter, an evaluation of SPR compensations is conducted on a Catphan phantom and an anthropomorphic chest phantom, and this new property is also directly applied to scatter estimation in CT imaging using a spectral modulator with flying focal spot technology (SMFFS). Preliminary results from our simulation as well as phantom studies show the effectiveness and potential application of the new property in practice.

The paper is organized as follows. In Section~\ref{sec:Method}, the scatter property is first developed and then applied to CT imaging using the SMFFS. After that, the experimental results are demonstrated in Section~\ref{sec:Results}. Finally, a brief discussion is given in Section~\ref{sec:Discussions} followed by a conclusion in Section \ref{sec:Conclusions}.

\section{Methods}\label{sec:Method}
\subsection{Linearization of Polychromatic Projection}
In X-ray CT, the detected X-ray intensity $I_t$ after passing through an object, in total, is the summation of primary after object attenuation, $I_p$, and scatter, $I_s$, i.e.,
\begin{linenomath}
\begin{equation}\label{eq_It}
  I_t=I_p+I_s=I_0\int{S(E) e^{-\mu(E)L} \mathrm{d}E}+I_s.
\end{equation} 
\end{linenomath}
where, $I_0$ is the initial X-ray intensity without object in the beam; $S(E)$ is a normalized effective spectrum of a polychromatic CT system ($\int{S(E)\mathrm{d}E}=1$); and $\mu(E)$ and $L$ are the linear attenuation coefficients of the object and its path-length, respectively. Here, for simplicity, a single-material object is assumed. Consequently, the total projection with scatter contamination can be written as:
\begin{linenomath}
\begin{equation}\label{eq_Pt}
    p_t = -\ln\left(\frac{I_p+I_s}{I_0}\right)=-\ln \left(\frac{\int {S(E)e^{-\mu (E)L} \mathrm{d}E}}{\int {S(E) \mathrm{d}E}}\right)-\ln {(1+{\rm SPR})}. 
\end{equation}
\end{linenomath}
Here, $\textrm{SPR}=I_s/I_p$ denotes the scatter-to-primary ratio (SPR).
According to Eq.~(\ref{eq_Pt}), a polychromatic projection free from scatter ($\textrm{SPR}=0$) can be defined by,
\begin{linenomath}
\begin{equation} \label{eq_poly_mono}
p=-\ln  \left({\frac {\int S(E)e^{-\mu^{'}(E) \cdot m} \mathrm{d}E}{\int S(E) \mathrm{d}E}}\right), 
\end{equation}
\end{linenomath}
with $\mu ^{'} (E) ={\mu (E)}/{\bar{\mu}}$ being the normalized attenuation coefficients, and $m =\bar{\mu}L$ being the corresponding monochromatic projection whose effective attenuation coefficient is set at $\bar{\mu}$.

Spectral correction for a polychromatic CT system, in essence, is to find a linearization mapping function, $f$, such that 
\begin{linenomath}
\begin{equation}
m=f(p), 
\end{equation}
\end{linenomath}
for each and every CT detector pixel in the scan field of view. The mapping, $f$, is usually called the beam hardening correction (BHC) function. For a single-material object, Eq.~(\ref{eq_poly_mono}) gives the inverse function of BHC function, i.e., $p=f^{-1}(m)$. By simulating or measuring polychromatic projections ($p$) at a range of object thicknesses ($L^{'}s$), one can build a linearization mapping function from polychromatic projections to their corresponding monochromatic ones.

However, with scatter uncorrected, the outcome of the linearization mapping 
is to applying BHC function, $f$, to the total projection, $p_t$, rather than the primary projection, $p$, causing errors in the corrected projection and CT image after reconstruction.

\subsection{The Characteristics of Scatter After Spectral Correction}

First, we want to characterize the scatter property to be presented as follows. After applying the spectral correction before scatter is removed,
\begin{itemize}
  \item \textbf{in the projection domain}: \textit{the impact of SPR on the CT projection will be scaled roughly by the first derivative of the linearization mapping function};
  \item \textbf{in the transmission domain}: \textit{the scatter distribution will be scaled roughly by the product of the first derivative of the mapping function and a natural exponential of the projection difference before and after the mapping}.
\end{itemize}
Detailed derivations are then given below.

\begin{figure}[htb]
	\centering
	\includegraphics[width=15cm]{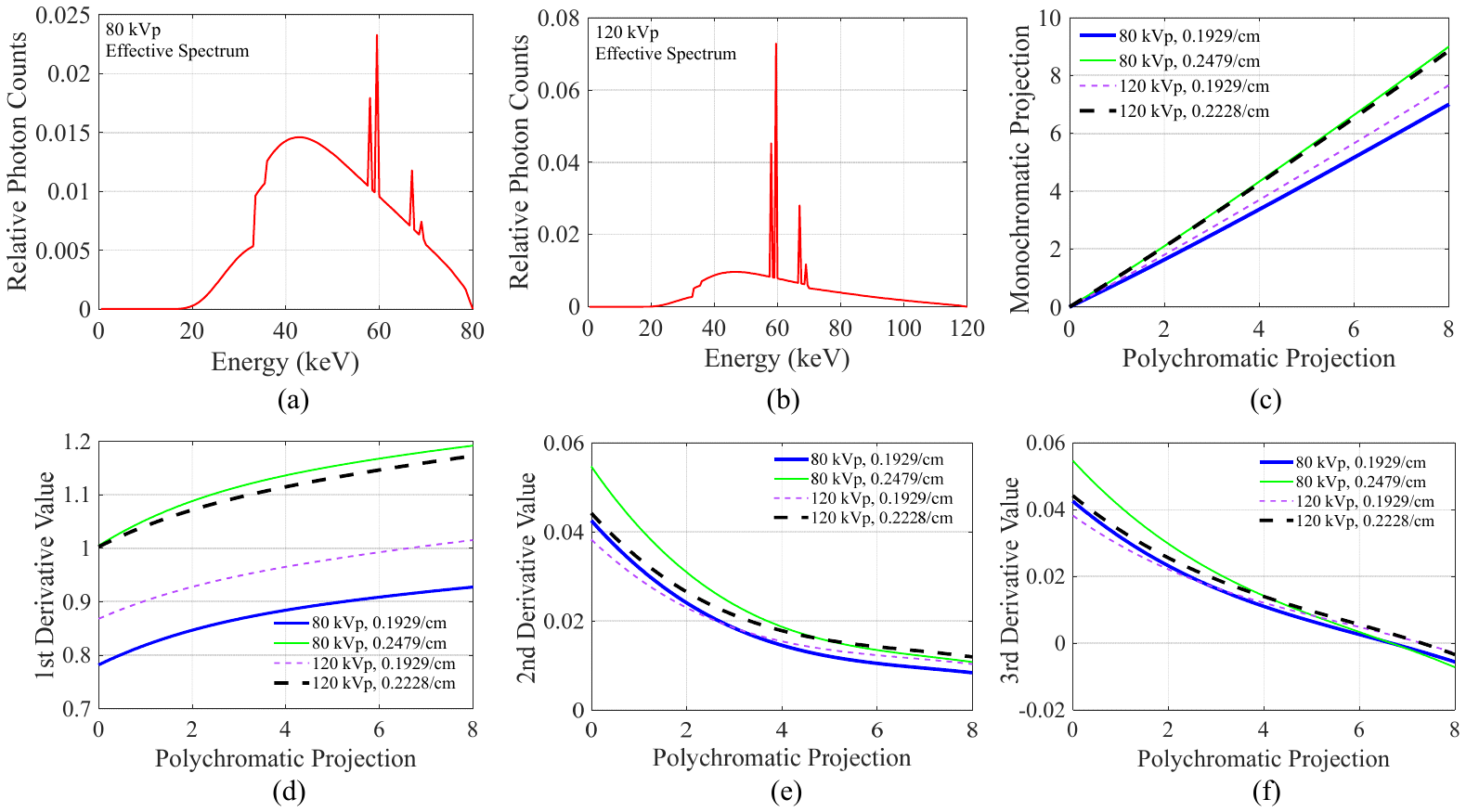}
	\caption[]{Beam hardening correction curves (BHC) and its derivatives: (a) and (b) the effective spectrum of a cone-beam CT system at
		80 kVp and 120 kVp, respectively; (c) the beam hardening curves of water for 80 kVp and 120 kVp effective spectrum using 
		different effective attenuation coefficient of water; (d) the first derivatives of water BHC curves in (c); 
		(e) the second derivatives of water BHC curves in (c);
		and (f) the third derivatives of water BHC curves in (c).\label{fig_BHC_Curves_Derivates}}
\end{figure}

Referring to Eq. (\ref{eq_Pt}), by using the Taylor expansion, the linearization mapping of total projection $p_t$ can be approximated as,
\begin{linenomath}
\begin{equation} \label{eq:proj_total_Taylor} 
f(p_t) =f(p-\Delta{p})=f(p)-f^{'}(p) \Delta{p}+\frac {1}{2} f^{''}(p) (\Delta{p})^2 \cdots ,
\end{equation}
\end{linenomath}
with $\Delta{p}=\ln {(1+SPR)}$. With Eq.~(\ref{eq_poly_mono}), one can easily derive the derivative of $f(p)$ for arbitrary orders.
Here, we explicitly give out the first three orders as follows\citep{gao2006beam}, with the first derivative being
\begin{linenomath}
\begin{equation} \label{eq_BHC_1st}
f^{'}(p)=\frac {\int S^{'}(E) \mathrm{d}E}{\int S^{'}(E) \mu^{'}(E) \mathrm{d}E},
\end{equation}
\end{linenomath}
where, $S^{'}(E)=S(E)e^{-\mu^{'}(E) \cdot f(p)}$; the second derivative being
\begin{linenomath}
\begin{equation} \label{eq_BHC_2nd}
f^{''}(p) 
=\left(\frac {\int S^{'}(E)  [\mu^{'}(E)]^2 \mathrm{d}E
  \times	
  \int S^{'}(E)\mathrm{d}E}{[\int S^{'}(E) \mu^{'}(E) \mathrm{d}E]^2}-1\right) f^{'}(p);
\end{equation}
\end{linenomath}
and the third derivative being
\begin{linenomath}
\begin{align} \label{eq_BHC_3rd}
                &f^{'''}(p)=\left(3\frac {\int S^{'}(E)  [\mu^{'}(E)]^2 \mathrm{d}E \times	\int S^{'}(E)\mathrm{d}E}{[\int S^{'}(E) \mu^{'}(E) \mathrm{d}E]^2}-1\right) f^{''}(p) \nonumber\\
                &+\left(\frac {\int S^{'}(E)  [\mu^{'}(E)]^2 \mathrm{d}E}{\int S^{'}(E) \mathrm{d}E}-\frac {\int S^{'}(E)  [\mu^{'}(E)]^3 \mathrm{d}E}{\int S^{'}(E) \mu^{'}(E) \mathrm{d}E}\right) [f^{'}(p)]^3.       
\end{align}  
\end{linenomath}
Due to the subtractions within Eqs.~(\ref{eq_BHC_2nd}) and (\ref{eq_BHC_3rd}), the second and third order derivatives of beam hardening correction function are both relatively small. For a better understanding, we plot the beam hardening correction curves of water at 80 kVp and 120 kVp spectra and their corresponding derivatives, respectively. To illustrate the influence of the effective water attenuation coefficient, $\bar{\mu}$, on the BHC curves, we select $\bar{\mu}$ to be the attenuation coefficient of water at 70 keV (0.1929 cm$^{-1}$), and the spectrum weighted attenuation coefficients [$\int S(E) u(E) {\rm d} E$, i.e., 0.2479 cm$^{-1}$ at 80 kVp, and 0.2228 cm$^{-1}$ at 120 kVp]. As shown in Fig.~\ref{fig_BHC_Curves_Derivates}, the second and third derivative values are both relatively small for all cases.

In addition, in practice most SPR (or the residual SPR after first-pass correction) for medical CT is less than 200\%, making $\Delta p=\ln (1+\textrm{SPR})<1$. Therefore, spectral correction for total projections with scatter contamination can be reasonably approximated by,

\begin{linenomath}
  \begin{align}\label{eq_BHC_apprx}
    f(p_t) &\approx f(p)-f^{'}(p) \ln{(1+SPR)} \nonumber \\
           &\approx f(p)-f^{'}(p_{t}) \ln{(1+SPR)}
 \end{align}     
\end{linenomath}
It is seen that the scatter caused discrepancy between primary projection $p$ and total projection $p_t$ is $\ln (1+{\rm SPR})$, which after spectral correction will be scaled approximately by the first derivative of the mapping function, $f^{'}(p)$. Given that the first derivative changes relatively slowly, $f^{'}(p)$ can be replaced by $f^{'}(p_{t})$, which is more convenient to obtain in practice. So far, we have obtained the first part of the scatter property in the beginning of this section.

Furthermore, by going back to the transmission domain, one can also figure out how the scatter distribution will change.
According to Eq.~(\ref{eq_BHC_apprx}), after spectral correction the total X-ray intensity can be written as,
\begin{linenomath}
\begin{align}\label{eq_BHC_approex_trans}
I^{'}_{t} &=I_0 e^{-f(p_t)}  \nonumber\\
&\approx I_0 e^{-f(p)} \cdot (1+SPR)^{f^{'}(p_{t})}.
\end{align}   
\end{linenomath} 
When taking ${f^{'}(p_{t})}\approx1$ into Eq.~(\ref{eq_BHC_approex_trans}) or $\textrm{SPR}$ is a small number, $I^{'}_{t}$ can be further simplified as,
\begin{linenomath}
\begin{align} \label{eq:I_c_t}
I^{'}_{t} &\approx I_0 e^{-f(p)}+I_{0} e^{-f(p)} \cdot f^{'}(p_{t}) SPR \nonumber\\
        &= I_0 e^{-f(p)}+I_{s}f^{'}(p_{t})e^{-f(p)+p}  \nonumber\\
        &\approx I_0 e^{-f(p)}+I_{s}f^{'}(p_{t})e^{-f(p_{t})+p_{t}} 
\end{align}
\end{linenomath}
Here, we use the facts that $\textrm{SPR}=I_s/I_p$ and $I_p=I_0e^{-p}$. It is also assumed that the projection difference before and after spectral correction, $p-f(p)$, can be approximated by, $p_{t}-f(p_{t})$, which is again more convenient to obtain in practice. Therefore, the new scatter distribution after spectral correction can be estimated by
\begin{linenomath}
\begin{equation} \label{eq:I_c_s}
  I_s^{'} \approx I_{s}f^{'}(p_{t})e^{[p_{t}-f(p_{t})]}.
\end{equation}
\end{linenomath}
It is seen that the scatter distribution after spectral correction is 
scaled approximately by a factor of $f^{'}(p_{t})e^{p_{t}-f(p_{t})}$, which is the product of the first derivative of the linearization mapping function and a natural exponential of the projection difference before and after the mapping.
Therefore, we have obtained the second part of the scatter property in the beginning of this section.

\subsection{Application of the Proposed Scatter Property} \label{subsec:application}

First, we note that in case the SPR is somehow available, after applying the spectral correction before scatter being removed, an SPR compensation (SPRC) can be done to recover $f(p)$ from $f(p_{t})$, in the following two ways.

 \begin{itemize}
   \item[1)] SPRC-I: $f(p) \approx f(p_t)+\ln{(1+SPR)}$, which totally ignores the impact of spectral correction;

   \item[2)] SPRC-II: $f(p) \approx f(p_t)+f^{'}(p_t)\ln{(1+SPR)}$, which uses $f^{'}(p_t)$ to adjust the SPR change due to spectral correction. 
 \end{itemize}
according to our proposed scatter property, SPRC-II will be more accurate. As a result, such an SPR compensation can be used to validate our proposed new scatter property indirectly.

Now, we want to apply the proposed new scatter property to CT imaging using a spectral modulator with flying focal spot (SMFFS) as an example to demonstrate its potential benefits in practical applications. SMFFS is a promising low-cost approach to accurately solving the X-ray scattering problem and physically enabling multi-energy imaging in a unified framework, with no significant sparsity or misalignment in data sampling\citep{gao2020spectral}. However, due to the X-ray spectral change from spectral modulation, correcting for scatter using SMFFS needs careful consideration and its related algorithm highly depends on the design and implementation of SMFFS. 

For SMFFS, when a focal spot $\textrm{FS}_k$ whose position is labeled as $k$ fires, the measured X-ray signal at a detector element $(i,j)$ can be expressed as:
\begin{linenomath}
\begin{equation}
I_{t_{k}}(i,j) = I_{p_k}(i,j)+I_{s_k}(i,j), \label{Eq_two_ray_a}       
\end{equation}
\end{linenomath}
where, $I_{p_k}(i,j)$ and $I_{s_k}(i,j)$ denote the intensities of the primary beam and the scatter, respectively, at source position $k$. 
And $I_{p_k}(i,j)$, can be further written as:
\begin{equation}
        I_{p_k}(i,j) = I_{m_k}\int S_{k}(E,i,j)e^{-u_{o}(E)L_{k}(i,j)}{\rm d} E. \label{Eq_Ip}
\end{equation}
Here, $I_{m_k}$ is the intensity of X-ray after beam passing through the modulator; $S_{k}(E,i,j)$ is the normalized spectrum with the modulator in the beam; $u_{o}(E)$ and $L_{k}(i,j)$ are the linear attenuation coefficients and the path-length of projection ray of the scanned object.

As investigated in the literature, scatter is mostly dominated by low-frequency components \citep{zhu2006scatter,gao2010scatter}. In SMFFS, it is further assumed that there is a strong correlation between scatter distributions at varying focal spot positions when the flying distance is relatively small \citep{gao2020spectral}. Based on our experience, it is also safe to assume that the low- and high-energy projection rays are aligned quite well (i.e., $L_1(i,j)=L_2(i,j)=L(i,j)$), and their scatter share the same distribution (i.e., $I_{s_{1}}(i,j)=I_{s_{2}}(i,j)=I_{s}(i,j)$). For simplicity, we omit the expression $(i,j)$ from the variables, e.g., $I_{t_{k}}$ represents $I_{t_{k}}(i,j)$.

For the two-point flying focal spot scenario in SMFFS, using the projection ray alignment and the scatter similarity, Eqs.~(\ref{Eq_two_ray_a}) and (\ref{Eq_Ip}) can be written as,
\begin{linenomath}
  \begin{subequations}\label{Eq_SMFFS_math}
    \begin{align} 
        I_{t_1} = I_{m_1}\int S_{1}(E)e^{-u_{o}(E)L}{\rm d} E+I_s, \label{Eq_SMFFS_math_1}\\ 
        I_{t_2} = I_{m_2}\int S_{2}(E)e^{-u_{o}(E)L}{\rm d} E+I_s.\label{Eq_SMFFS_math_2}
   \end{align}
  \end{subequations}
\end{linenomath}

To estimate the scatter distributions for SMFFS, we need to the solve the unknown variable $I_s$ in Eqs.~(\ref{Eq_SMFFS_math}). 
A scatter estimation method for SMFFS based on our proposed scatter property
is given in the following. 

For the data acquired from SMFFS, after spectral correction one can write the total X-ray intensity as:
\begin{linenomath}
\begin{subequations}\label{eq:SMFFS}
  \begin{align} 
    I_{t_1}^{'}=I_{m_1}e^{-f_1(p_{t_1})} = I_{m_1}e^{-f_1(p_1)}+I_{s_1}^{'},\\
  I_{t_2}^{'}=I_{m_2}e^{-f_2(p_{t_2})} = I_{m_2}e^{-f_2(p_2)}+I_{s_2}^{'},
\end{align}
\end{subequations}
\end{linenomath}
where, $p_{t_{k}}=-\ln (I_{t_{k}}/I_{m_{k}})$ is the total projection with scatter; $p_{k}=-\ln (I_{p_{k}}/I_{m_{k}})$ is the projection without scatter, $f_k$ is the spectral correction linearization mapping function and $f'_k$ is the first order derivative of $f_k$. 

So, if one simply ignores the impact of spectral correction on the distribution of scatter (i.e., $I_s^{'}=I_s$), then $I_s$ can be easily computed using Eq.~(\ref{eq:SMFFS}) as
\begin{linenomath}
\begin{equation}\label{eq:I_SE_D}
  I_s\approx\frac{I_{m_1}I_{m_2}\left[e^{-f_1(p_{t_1})}-e^{-f_2(p_{t_2}})\right]}{I_{m_2}-I_{m_1}}.
\end{equation}
\end{linenomath}
This oversimplified scatter estimate is usually biased and less accurate in practice, which is denoted as ${\rm SE_d}$.

According to our proposed scatter property, however, we know that the scatter distribution after spectral correction will be scaled by the product of the first derivative of the linearization mapping function and a natural exponential of the projection difference before and after the mapping. So, by substituting $I_{s_k}^{'}$ in Eq.~(\ref{eq:I_c_s}) into Eq.~(\ref{eq:SMFFS}), Eq.~(\ref{eq:I_SE_D}) should be revised as
\begin{linenomath}
\begin{equation}\label{eq:I_SE_SPS}
  I_s\approx\frac{I_{m_1}I_{m_2}\left[e^{-f_1(p_{t_1})}-e^{-f_2(p_{t_2}})\right]}{I_{m_2}f^{'}_{1}\left(p_{t_{1}}\right)e^{[p_{t_1}-f_1(p_{t_1})]}-I_{m_1}f^{'}_{2}\left(p_{t_{2}}\right)e^{[p_{t_2}-f_2(p_{t_2})]}}.
\end{equation}
\end{linenomath}
Hence, an analytic form of an accurate scatter estimate in this paper has been obtained with no iteration needed, which is denoted as ${\rm SE_{sps}}$. 

For the SMFFS and other similar scenarios, it is worth noting that due to the spatially-varying modulator in the beam, the effective X-ray spectrum at each detector pixel could be quite different from one to another. In other words, the spectral correction mapping functions, $f$, across the entire scan field of view vary spatially too\citep{gao2019physics}. As a result, after spectral correction the new scatter distribution will be significantly distorted (according to our proposed scatter property) and the assumption of a low-frequency distribution may no longer be appropriate. 

There exist other ways to estimate the scatter for SMFFS using relations in Eqs. (\ref{Eq_SMFFS_math}). For comparison with the proposed method ${\rm SE_{sps}}$ , in Appendix A we also give another scatter estimation method for SMFFS based on residual spectral linearization (denoted as ${\rm SE_{rsl}}$).

\section{Evaluations and Results}\label{sec:Results}
\subsection{The Experimental Setups}
\begin{table}[htb]
  \centering
  \caption[]{Parameters Selection in the Simulation Study.}\label{table_simulation_study}
 {
  \begin{tabular}{ll}
  \toprule
  Parameters & Value range  \\  
  \midrule
  Thickness of molybdenum ($T_{Mo}$) & 0.2 $\sim$ 0.8 mm\\
  Thickness of copper ($T_{Cu}$) & 0.2 $\sim$ 0.8 mm\\
  Water fraction ($\beta$) & 0.3 $\sim$ 1\\                          
  Scatter deviation ($\alpha$) & -0.2 $\sim$ 0.2 \\
  \bottomrule
\end{tabular}
}
\end{table}

First, a simulation study of estimated scatter error was conducted to compare the performances 
of the methods $\rm{SE_{rsl}}$ and $\rm{SE_{sps}}$, where the low- and high-energy X-rays passing through the same path of a object was simulated, with parameters selection given in Table \ref{table_simulation_study}. The details about the simulation experiments will be described in section \ref{sec:Data_acquire}

\begin{table}[htb]
  \centering
  \caption[]{Imaging Parameters of a Tabletop CBCT System.}\label{table_geometry}
  \resizebox{12cm}{!}{
  \begin{tabular}{lll}
  \toprule
  &Parameters & Values\\  
  \midrule
  \multirow{7}{*}{Geometry} &Focal spot & 0.3 mm \\
  & Source-detector distance & 1419.6 mm\\
  &Source-isocenter distance & 860.9 mm\\ 
  &\multirow{2}{*}{Detector size} & 0.39$\times$0.39 mm$^2$\\ 
  & &1024$\times$768 pixels \\                             
  &Scan & circular, 360 degrees\\
  &Number of projection views  & 625 (No FFS); 1250 (FFS)\\
  \cmidrule(lr){1-3}
  \multirow{4}{*}{SMFFS} &Modulator material & copper \\
  &Modulator thickness & 0.42 mm\\
  &Modulator spacing &0.889 mm\\
  &Focal-spot deflection distance &1.24 mm\\
  \bottomrule
\end{tabular}
}
\end{table}

In addition, we further validated our developed scatter property on a tabletop CBCT system with some major imaging parameters listed in Table \ref{table_geometry}. The X-ray tube was operated at 80 kVp and 120 kVp, and the effective spectra were obtained by an expectation method \citep{duan2011ct} as shown in Fig. \ref{fig_BHC_Curves_Derivates} (a) and (b). To conduct CT scans with SMFFS, a stationary copper filter in a 2D checkboard pattern with 0.42 mm thickness and 0.889 mm spacing was used as the spectral modulator and the X-ray tube was moved by 1.24 mm in the X direction to mimic the flying focal spot. An adaptive spectral estimation and compensation algorithm in Ref \citep{gao2019physics} was adopted to capture the overall detected spectrum of each and every detector pixels in the presence of the spectral modulator across the scan field of view at each of the focal spot positions, respectively.

\subsection{CT Data Acquisition and Processing} \label{sec:Data_acquire}
In this work, to validate our proposed scatter property, we conducted in-depth evaluations of the SPR compensation methods, a simulation study of estimated scatter error and the performance of scatter correction from CT scans using the SMFFS. 

\begin{figure}
  \centering
  \includegraphics[width=13cm]{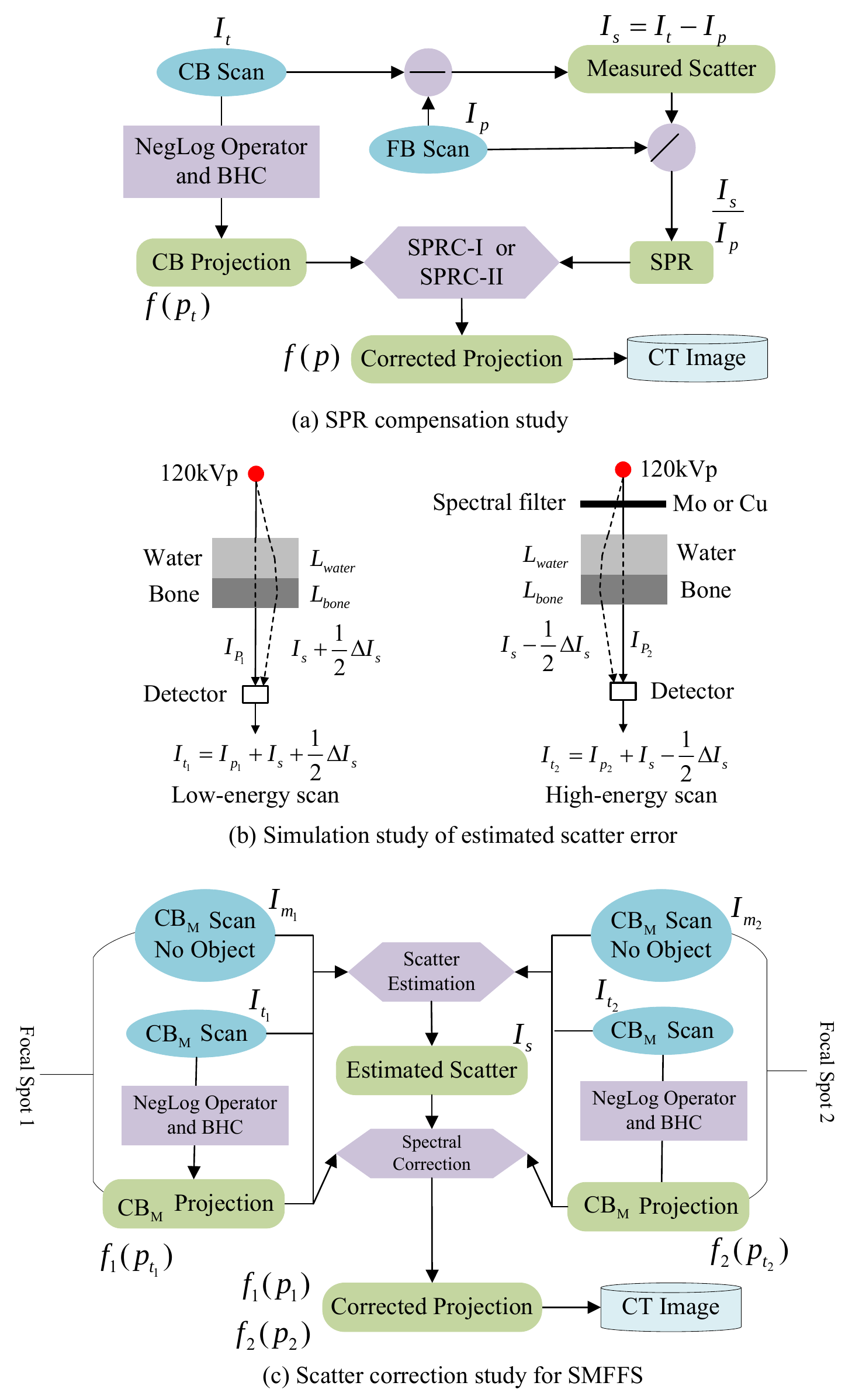}
  \caption[]{Data processing flowcharts for a SPR compensation study, a simulation study of estimated scatter error and an evaluation for scatter correction on a SMFFS system. 
  \label{fig_Data_Process}}  
\end{figure}

For the SPR compensation, evaluations were performed using a Catphan phantom and an anthropomorphic chest phantom. As illustrated in Fig.~\ref{fig_Data_Process}(a), we took a fan-beam (FB) scan and a cone-beam (CB) scan of the phantoms, with an assumption that the fan-beam scan is scatter free. Thus, scatter distributions can be obtained from subtracting the fan-beam transmission image from the cone-beam transmission image in the transmission domain, and then SPR was computed by the ratio of the measured scatter to the fan-beam transmission image. Finally, projections for reconstructing CT images were recovered using the SPR compensation method as described in Section~\ref{subsec:application}.

For the simulation study of estimated scatter error, the low- and high-energy X-rays passing through the same path of a object was simulated. As shown in Fig. \ref{fig_Data_Process}(b), the low-energy X-ray was generated by the simulated 120 kVp spectrum while the high-energy was obtained after filtering the 120 kVp spectrum with molybdenum (thickness denoted by $T_{Mo}$) or copper (thickness denoted by $T_{Cu}$). And the object was simulated by a combination of water and bone, with the length being $L_{water}$ and $L_{bone}$ for water and bone, respectively.
We define the water fraction as,

\begin{equation}
  \beta = \frac{\overline{u}_{water}L_{water}}{\overline{u}_{water}L_{water}+\overline{u}_{bone}L_{bone}},
\end{equation}
here, $\overline{u}_{water}$ and $\overline{u}_{bone}$ are the effective linear attenuation coefficients of water and bone for the spectrum at 120kVp, which are 0.2228 cm$^{-1}$ and 0.7649 cm$^{-1}$, respectively. A larger $\beta$
indicates that the relative contribution from bone is less in the corresponding object path-length. 
The primaries after the object attenuation were denoted as $I_{p_1}$ and $I_{p_2}$ for the low- and high-energy projection rays.
Considering that the assumption of scatter similarity for low- and high-energy projection rays may not hold strictly (i.e., $I_{s_1}\neq I_{s_2}$), the simulated scatter 
consisted of nominal part ($I_s$) and variation part ($\Delta I_s=I_s*\alpha$, with $\alpha$ being the scatter deviation). The detected X-ray total intensity was given by summing up the scatter and primary after the object attenuation. As a result, the detected X-ray total intensity of high- and low-energy projection rays are $I_{t_1}=I_{p_1}+I_s+\frac{1}{2}\Delta I_s$ and $I_{t_2}=I_{p_2}+I_s-\frac{1}{2}\Delta I_s$, respectively.
The parameters $T_{Mo}$, $T_{Cu}$, $\beta$ and $\alpha$ can be adjusted in order to
simulate a range of imaging cases.
Finally, a comparison analysis between the methods $\rm{SE_{rsl}}$ and $\rm{SE_{sps}}$ was carried out on the error of the estimated scatter of the simulated data. 

For the scatter correction from CT scans using the SMFFS, data acquisition and processing are illustrated in Fig.~\ref{fig_Data_Process}(b). Cone-beam scans with the modulator (${\rm CB_M}$) scans were taken with and without the Catphan phantom, with the X-ray source operated at 120 kVp. Scatter distributions are  estimated using Eq.~(\ref{eq:I_SE_D}), Eq.~(\ref{Eq_SMFFS_scatter_rsl}) and Eq.~(\ref{eq:I_SE_SPS}), respectively. 
And projections after scatter correction and spectral correction are obtained and ready for reconstruction. 

The CT images were reconstructed using the standard Feldkamp-Davis-Kress (FDK) algorithm \citep{feldkamp1984practical}. The reconstructed images were presented in Hounsfield units (HU), with a CT number of -1000 HU for air and a CT number of 0 HU for water-equivalent materials. To quantify the performance, we calculated the averaged CT number for the selected regions of interest (ROIs), and measured the root mean square error (RMSE) defined as

\begin{equation}\label{eq:RMSE}
E_{{\rm RMSE}} = \sqrt{\frac{1}{N}\sum_{n=1}^{N}\left(\mu_{n}-\widetilde{{u_{n}}}\right)^2},
\end{equation}

where, $n$ is the index of ROIs, $N$ is the total number of ROIs; $\mu_{n}$ is the averaged  CT number inside ROI $n$ of the reconstructed CT image; and $\widetilde{\mu_{n}}$ is the corresponding value of the reference image. In this study, the fan-beam CT images after spectral correction are taken as the ground-truth reference.

To quantitatively evaluate the scatter estimation performance in the simulation study, a scatter error was also calculated as follows:

\begin{equation}\label{eq:Error_scatter}
Err_{\textrm{scatter}} =  \frac{\lvert \hat{ I_{s}}-\widetilde{{{I_{s}}}} \rvert}{\lvert\widetilde{I_{s}}\rvert} 
\end{equation}
where $\hat{ I_{s}}$ and $\widetilde{I_{s}}$ are the estimated and reference scatter values, respectively.

\subsection{SPR Compensation Study}
\begin{figure}
	\centering
	\includegraphics[width=10cm]{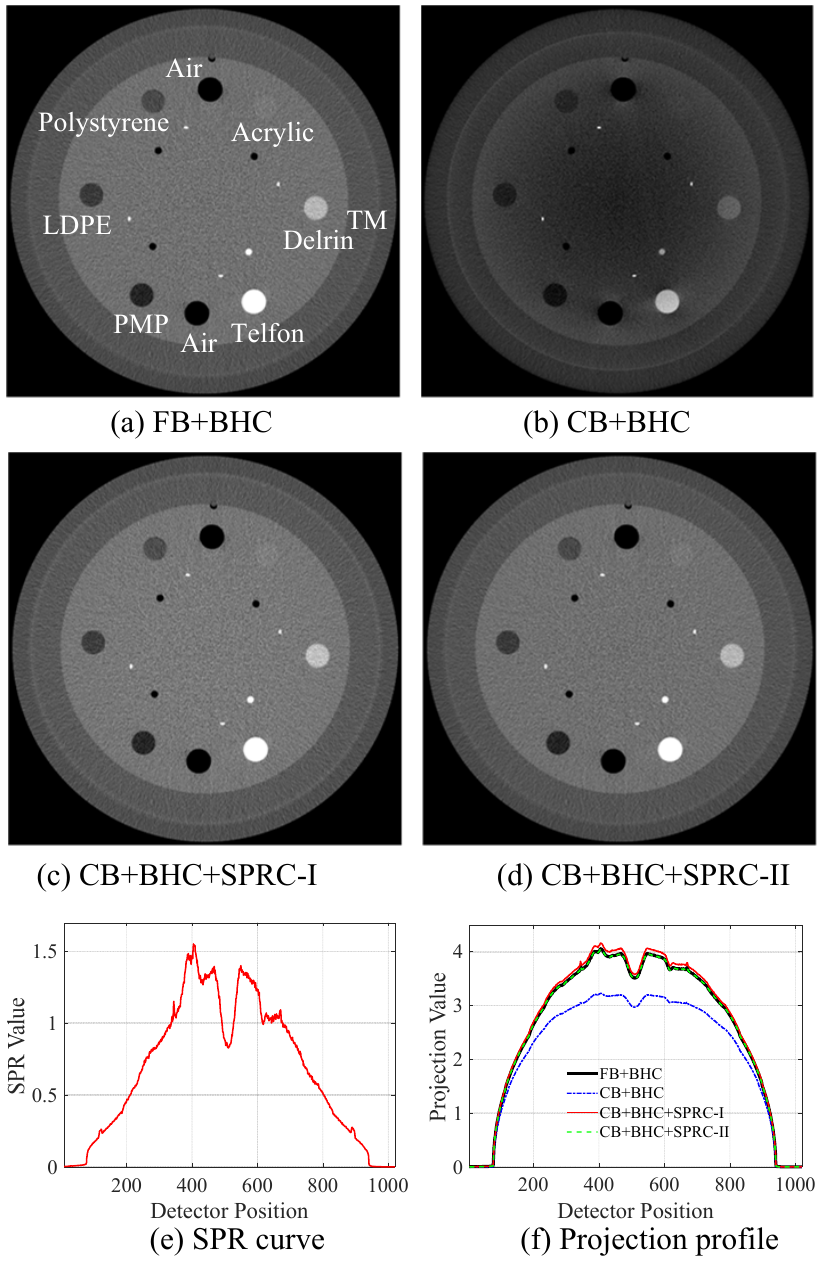}
	\caption[]{Reconstructed CT image from the fan-beam CT scan (reference) and cone-beam CT scan of the Catphan phantom at 80 kVp after spectral correction (beam hardening correction specifically) with $\bar{u}$=0.1929 cm$^{-1}$, with and without SPR compensation. Reconstruction size: $512\times512$ pixels with $0.4\times0.4$ mm$^2$ pixel size. (a) Fan-beam (FB) CT image with beam hardening correction (BHC); (b) cone-beam (CB) CT image with BHC; (c) cone-beam CT image with BHC followed by SPR compensation SPRC-I method; (d) cone-beam CT image with BHC followed by SPRC-II method; (e) SPR profile of the Catphan phantom; and (f) projection profiles of the fan-beam and cone-beam projections after BHC and followed by SPRC-I and SPRC-II, respectively. Display window: [-350 550] HU for (a)-(d).
		\label{fig_Catphan_SPR_Projection_Recon}  }
\end{figure}

For the Catphan phantom, the CT images reconstructed from fan-beam and cone-beam projection data with and without SPR compensation are demonstrated in Fig. \ref{fig_Catphan_SPR_Projection_Recon}, where the SPR and projection profiles are also provided as a comparison. Spectral correction (specifically beam hardening correction in this case as the spectra across the scan field of view is unchanged) was applied to all projection data. 
Please note that the ring artifacts near the edge of the Catphan phantom of fan-beam and cone-beam scans are due to an imperfect dynamic gain calibration on our flat panel detector.
Taking the fan-beam CT image as reference, cupping and shading artifacts caused by scatter are notably observed in the cone-beam CT image, which are significantly reduced by the SPR compensation methods: SPRC-I and SPRC-II, respectively. On the cone-beam CT projection after beam hardening correction, due to the uncorrected scatter there are obvious discrepancies when compared with the fan-beam projection. These discrepancies are narrowed by using both SPRC-I and SPRC-II. However, the projection profile by the SPRC-II method is closer to and superimposes better over the reference (fan-beam projection).

To quantify the performance of the SPRC-II method,  as listed in Table~\ref{table_Catphan_SPRC_ROI} we calculated the averaged  CT number in the selected ROIs indicated in Fig.~\ref{fig_Catphan_SPR_Projection_Recon}(a), as well as the RMSE defined in Eq.~(\ref{eq:RMSE}). It is seen that the accuracy of the averaged CT number within the selected ROIs is improved by using our proposed SPRC-II method, with the $E_{{\rm RMSE}}$ being reduced from 305.8 to 35.8 HU by SPRC-I, while further reduced to 2.9 HU by SPRC-II. 

\begin{table}[htb]
	\centering
	\caption[]{Averaged  CT number (HU) of selected ROIs of reconstructed CT images of the Catphan phantom in Fig.~\ref{fig_Catphan_SPR_Projection_Recon}.}\label{table_Catphan_SPRC_ROI}
	\resizebox{15cm}{!}{
		\begin{tabular}{ccccccccc}
			\toprule
			ROIs & Air & PMP & LDPE & Polystyrene & Acrylic & Delrin$^{\rm TM}$ & Teflon & $E_{{\rm RMSE}}$\\
			\midrule
			FB + BHC & -953.5 & -225.1 & -144.4 & -88.4 & 72.3 & 293.9 & 934.7 &  $\backslash$ \\
			CB + BHC& -696.0 & -292.2 & -245.4 & -218.3 & -132.3 & -18.2 & 288.7 & 305.8 \\
			\midrule
			CB + BHC + SPRC-I& -980.6 & -215.1 & -130.5 & -71.8 & 97.9 & 331.5 & 1009.4 & 35.8\\     
			CB + BHC + SPRC-II& -950.3 & -225 & -144.7 & -89 & 71.0 & 291.4 & 928.4 & 2.9\\     
			\bottomrule
		\end{tabular}
	}
\end{table}

\begin{figure}
	\centering
	\includegraphics[width=16cm]{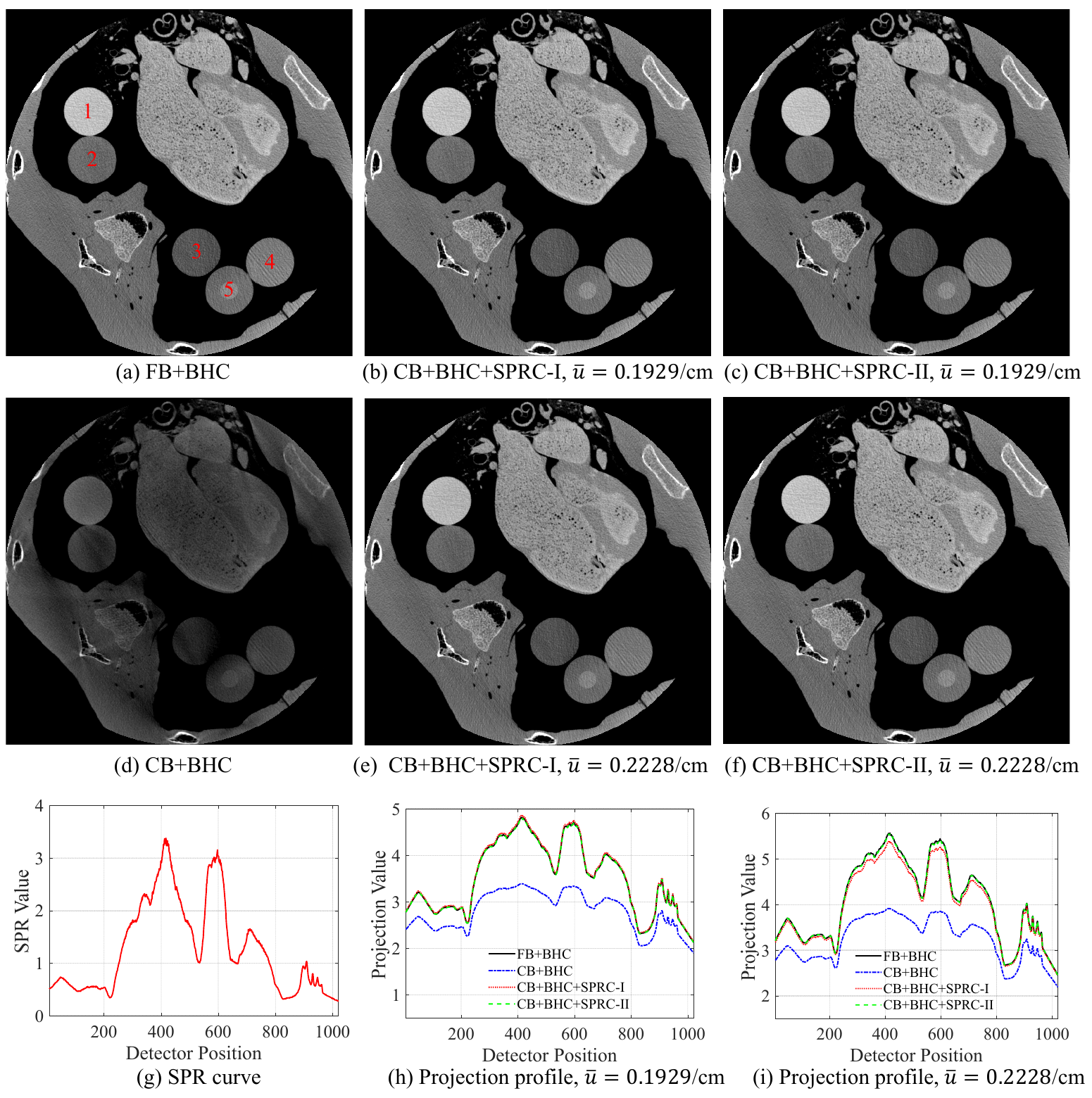}
	\caption{Reconstructed CT image from the fan-beam CT scan (reference) and cone-beam CT scan of the chest phantom at 120 kVp after spectral correction (beam hardening correction specifically) with and without SPR compensation, using $\bar{u}$=0.1929 cm$^{-1}$ and $\bar{u}$=0.2228 cm$^{-1}$, respectively. Reconstruction size: $512\times512$ pixels with $0.4\times0.4$ mm$^2$ pixel size. (a) Fan-beam (FB) and (d) cone-beam (CB) CT image with beam hardening correction (BHC); (b) and (e) cone-beam CT image with BHC followed by SPRC-I with $\bar{u}$=0.1929 cm$^{-1}$ and $\bar{u}$=0.2228 cm$^{-1}$, respectively; (c) and (f) cone-beam CT image with BHC followed by SPRC-II with $\bar{u}$=0.1929 cm$^{-1}$ and $\bar{u}$=0.2228 cm$^{-1}$, respectively; (g) SPR profile of the chest phantom; and (h) and (i) projection profiles of the fan-beam and cone-beam projections after BHC and followed by SPRC-II with $\bar{u}$=0.1929 cm$^{-1}$ and $\bar{u}$=0.2228 cm$^{-1}$, respectively. Display window: [-500 800] HU for (a)-(f).
	\label{fig_Chestphan_SPR_Projection_Recon_120kVp}  }
\end{figure}

We further evaluate our proposed characterization of scatter property on the anthropomorphic chest phantom. A comparison is also provided to show how the derived SPR compensation performs as the value of the effective attenuation coefficient $\bar{u}$ in the spectral correction (linearization mapping function) varies. Fig.~\ref{fig_Chestphan_SPR_Projection_Recon_120kVp} demonstrates CT images of the chest phantom reconstructed from fan-beam and cone-beam projection data after SPRC-I and SPRC-III compensation with $\bar{u}$=0.1929 cm$^{-1}$ and $\bar{u}$=0.2228 cm$^{-1}$, corresponding to the linear attenuation coefficients at 70 keV and the effective linear attenuation coefficient for the 120kVp spectrum, respectively. The SPR and projection profiles are also provided as a comparison. To quantitatively evaluate the image quality, five ROIs marked in Fig.~\ref{fig_Chestphan_SPR_Projection_Recon_120kVp}(a) are selected, whose averaged CT numbers are summarized in Table~\ref{table_Chestphan_SPRC_ROI}. Although the shape of SPR and its range of values on the chest phantom are more challenging, it is seen that our derived SPRC-II method again outperformed SPRC-I, with projection profiles closer to the reference. The $E_{{\rm RMSE}}$ error was reduced from 378.9 to 10.3 and 42.0 HU by SPRC-I with $\bar{u}$=0.1929 cm$^{-1}$ and $\bar{u}$=0.2228 cm$^{-1}$, respectively, while further reduced to 4.4 HU by SPRC-II for both $\bar{u}$. 
It is observed that selection of $\bar{u}$ does not change the outcome of SPRC-II, while
the performance of SPRC-I is dependent on $\bar{u}$ as expected. This is because the values of linearization mapping function and its first derivative are proportional to $\bar{u}$ while the term $\ln(1+SPR)$ is not, and the reconstructed result will be scaled by $\bar{u}$ to obtain the final CT number. 

\begin{table}[htb]
	\centering
	\caption[]{Averaged CT number (HU) of selected ROIs of reconstructed CT images of the chest phantom in Fig.~\ref{fig_Chestphan_SPR_Projection_Recon_120kVp}.}\label{table_Chestphan_SPRC_ROI}
	\resizebox{13cm}{!}{
		\begin{tabular}{lcccccc}
			\toprule
			ROIs & 1 & 2 & 3 & 4 & 5 & $E_{\rm RMSE}$\\
			\midrule
			FB + BHC& 481.1&36.0  & -64.9 &134.9  &202.4 & $\backslash$\\
			CB + BHC& -69.8 &-269.8  &-352.2  &-184.2  &-166.9  & 378.9\\
			\midrule
			CB+ BHC+SPRC-I, $\bar{u}$=0.1929  & 494.8&45.3&-56.6&144.2&212.3&10.3\\

			CB+ BHC+SPRC-I, $\bar{u}$=0.2228   &419.0&3.0&-96.2&100.1&161.4&42.0\\

			CB+ BHC+SPRC-II, $\bar{u}$=0.1929  &474.2 &32.9&-67.9&131.4&198.3&4.4\\

			CB+ BHC+SPRC-II, $\bar{u}$=0.2228  &474.2 &32.9&-67.9&131.4&198.3&4.4\\  
			
			\bottomrule
		\end{tabular}
	}
\end{table}
Based on the side-by-side comparison and RMSE analysis, the preliminary results on the Catphan and chest phantoms validate the effectiveness of the derived SPRC-II method and hence the correctness of our proposed scatter property. What can be also concluded is that the performance of the SPRC-II method is insensitive to the selection of $\bar{u}$. 

\subsection{Simulation Study of Estimated Scatter Error}
\begin{figure}[htb]
	\centering
	\includegraphics[width=16cm]{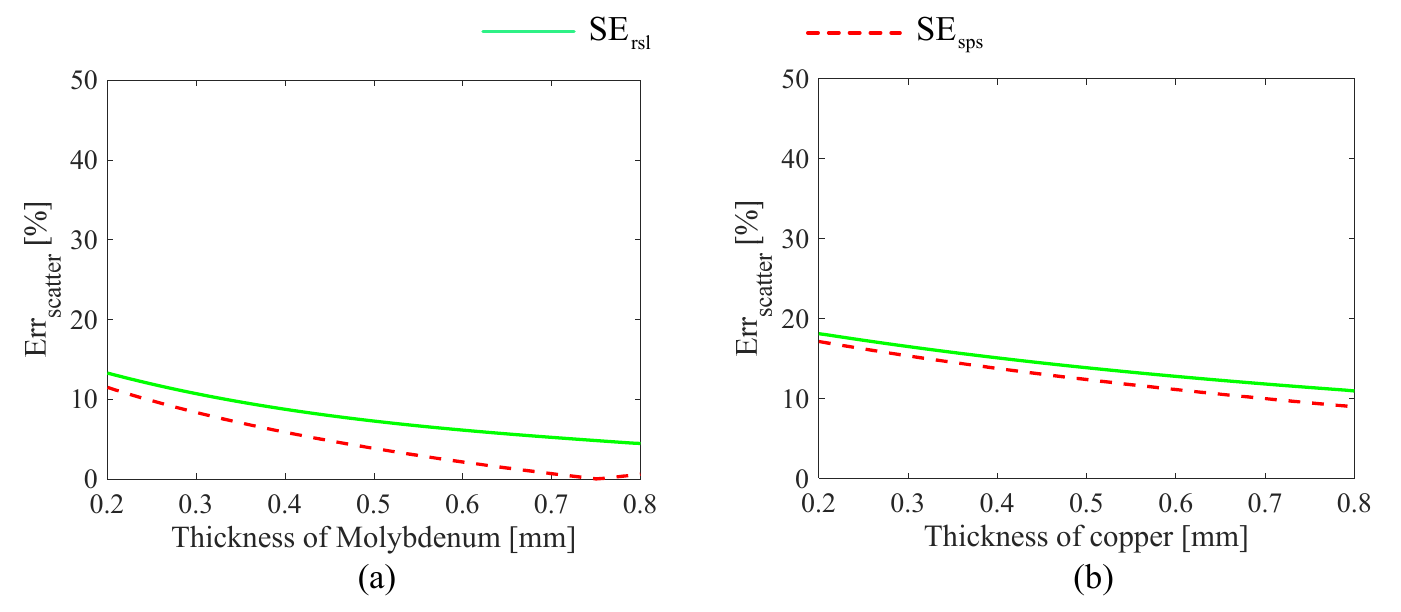}
  \caption[]{Error analysis of the estimated scatter from simulated data by the $\rm{SE_{rsl}}$ and $\rm{SE_{sps}}$ methods.
  (a) The spectral filter is molybdenum and the thickness of molybdenum varies from 0.2 to 0.8 mm, with the water fraction and the scatter deviation being 0.8 and 0, respectively;
  (b) the spectral filter is copper and the thickness of copper varies from 0.2 to 0.8 mm, with the water fraction and the scatter deviation being 0.8 and 0, respectively.
\label{fig_Simulation_Study}  }
\end{figure}

\begin{figure}
	\centering
	\includegraphics[width=15cm]{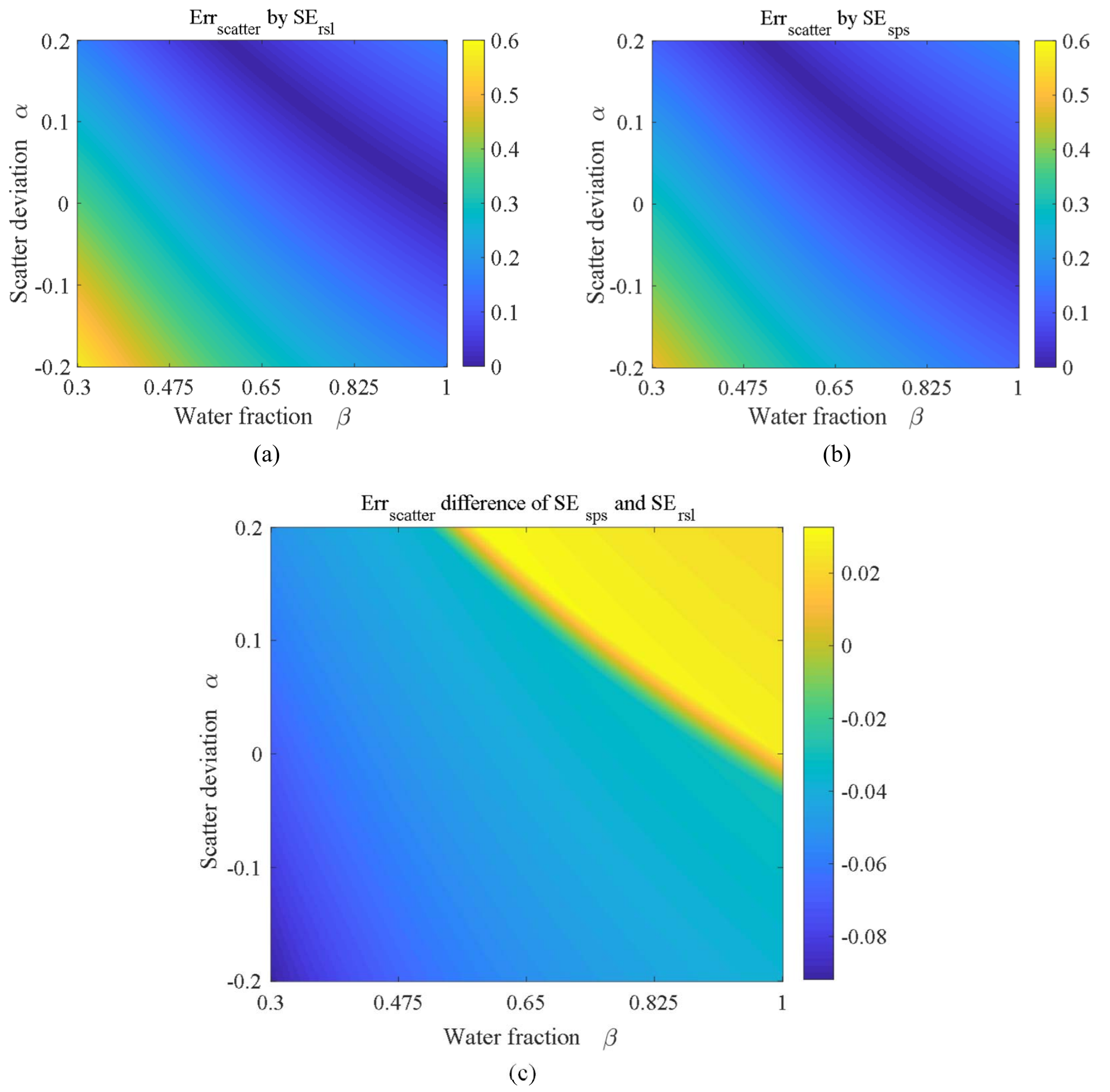}
  \caption[]{Scatter error comparison between the ${\rm SE_{rsl}}$ of ${\rm SE_{sps}}$ methods with water fraction and scatter deviation at various values.
  Here, the spectral filter was the molybdenum of a thickness being 0.5 mm.
  (a) Scatter error $Err_{\rm scatter}$ obtained by ${\rm SE_{rsl}}$ from the simulation data;
  (b) scatter error $Err_{\rm scatter}$ obtained by ${\rm SE_{sps}}$ from the simulation data;
  (c) the values of $Err_{\rm scatter}$ by ${\rm SE_{sps}}$ subtracting that by ${\rm SE_{rsl}}$.
\label{fig_Simulation_Study_v2}  }
\end{figure} 

In this section, we compared the scatter estimation results of the methods $\rm{SE_{rsl}}$ and $\rm{SE_{sps}}$ from the simulated data. 
The error $Err_{\rm scatter}$ defined in Eq. (\ref{eq:Error_scatter}) was calculated 
to measure the accuracy of scatter estimation. 
First, a range of thicknesses of molybdenum or copper were chosen to simulate different high-energy spectra and Fig. \ref{fig_Simulation_Study} demonstrates how $Err_{\rm scatter}$ by the ${\rm SE_{rsl}}$ and ${\rm SE_{sps}}$ methods varies with these thicknesses. It indicates that, overall, the scatter estimation performances of the methods $\rm{SE_{rsl}}$ and $\rm{SE_{sps}}$ are close. 
It is also seen that molybdenum performs better as the spectral filter material, relative to the copper. So, in the second part of our simulation study, only molybdenum is used. 
As shown in Fig. \ref{fig_Simulation_Study_v2}, the $Err_{\rm scatter}$ by the ${\rm SE_{rsl}}$ and ${\rm SE_{sps}}$ methods were calculated with the water fraction with the scatter deviation at different values with 
the spectral filter fixed to 0.5 mm thick molybdenum.   
In order to compare the $Err_{\rm scatter}$ by two methods clearly, we also calculated the 
values of $Err_{\rm scatter}$ of ${\rm SE_{sps}}$ subtracting that of ${\rm SE_{rsl}}$.
The difference results indicate that, with larger values of the water fraction and the scatter deviation, the $Err_{\rm scatter}$ of ${\rm SE_{rsl}}$ is smaller than that of ${\rm SE_{sps}}$. 
In the ideal case with the water fraction and scatter deviation being 1 and 0, respectively, the $Err_{\rm scatter}$ of ${\rm SE_{rsl}}$ will be zero which is smaller than that of ${\rm SE_{sps}}$ as expected. 
However, in the other cases, the ${\rm SE_{sps}}$ method performs slightly better than ${\rm SE_{rsl}}$ in terms of scatter estimation error. 
Therefore, it can be seen that even though ${\rm SE_{sps}}$ is an approximate form of scatter estimation based on our proposed scatter property, in practical application, its performance will not be inferior to ${\rm SE_{rsl}}$, which is based on a more strict derivation but requires an extra correction function from a residual spectrum.

\subsection{Evaluation of Scatter Correction from CT Scans using the SMFFS}
Here, we preliminarily evaluate the scatter correction algorithms using CT scans of the Catphan phantom with SMFFS. The ${\rm {SE_{d}}}$, ${\rm {SE_{rsl}}}$ and ${\rm {SE_{sps}}}$ algorithms were implemented to correct for scatter from a cone-beam scan with SMFFS. 

\begin{figure}[htb]
	\centering
	\includegraphics[width=15cm]{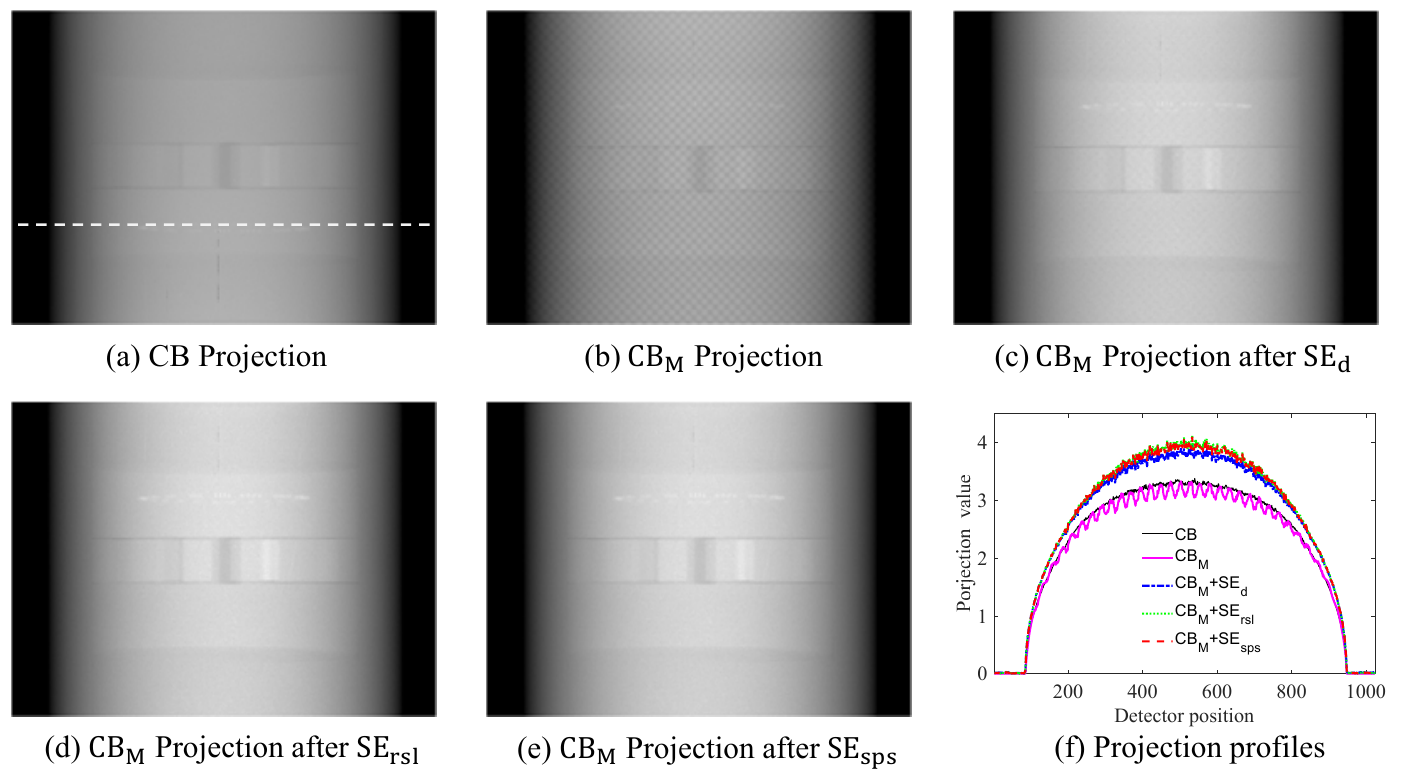}
  \caption[]{Projections of the Catphan phantom. (a) Projection from a cone-beam (CB) scan; (b) projection from a cone-beam scan with modulator (${\rm CB_M}$); (c) projection from a ${\rm CB_M}$ scan with ${\rm SE_d}$; (d) projection from a ${\rm CB_M}$ scan with ${\rm SE_{rsl}}$ 
  (e) projection from a ${\rm CB_M}$ scan with ${\rm SE_{sps}}$; and (f) the projection profiles of (a)$\sim$(e) along the line indicated in (a).
		\label{fig_Catphan_Scatter_Correction_Projection}  }
\end{figure}

\begin{figure}[htb]
	\centering
	\includegraphics[width=15cm]{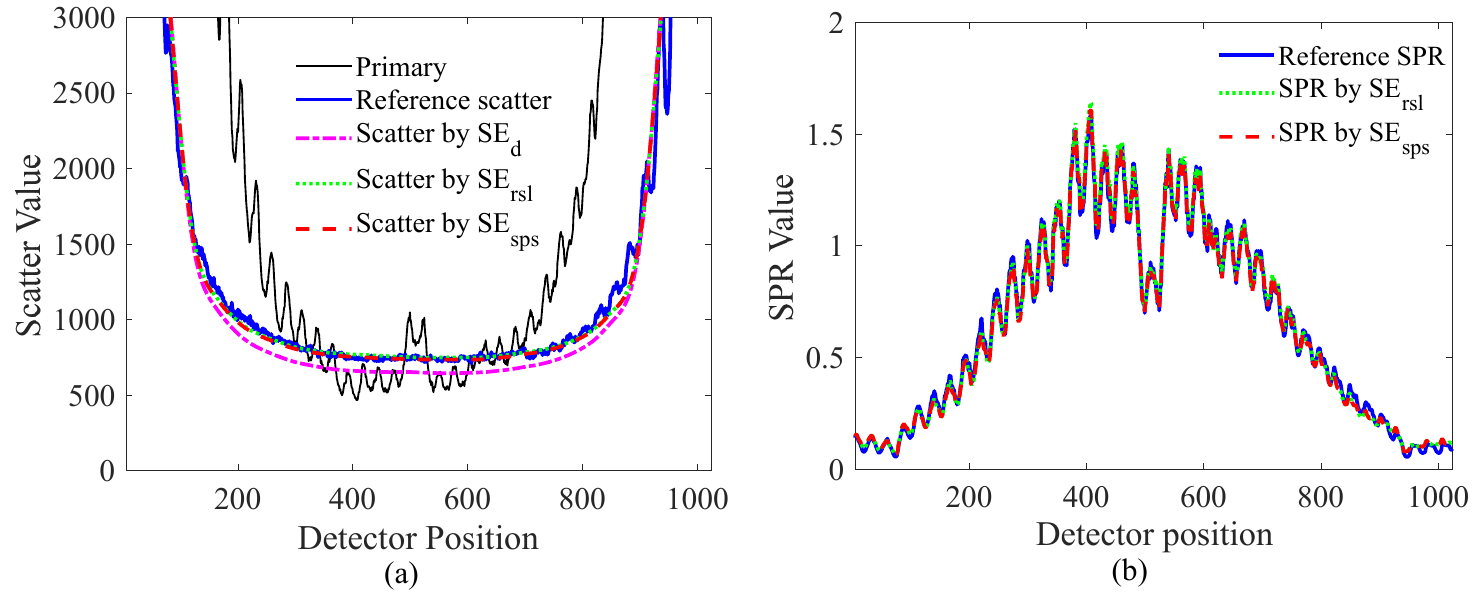}
  \caption[]{Scatter and SPR profiles of the cone-beam scan with modulator
  obtained by the ${\rm {SE_{d}}}$, ${\rm {SE_{rsl}}}$ and ${\rm {SE_{sps}}}$ methods, with an assumption that the fan-beam scan is free from scatter (i.e., primary signal only). (a) is the scatter profile and (b) is the SPR profile.
		\label{fig_Catphan_Scatter_Correction_Scatter}  }
\end{figure}

The projections of the Catphan phantom from the cone-beam scan with and without modulator at 120 kVp are shown in Fig.~\ref{fig_Catphan_Scatter_Correction_Projection}. 
Spectral correction was applied to all projection data. 
It is seen from the projection images that the modulator pattern can be better removed by the ${\rm SE_{rsl}}$ and ${\rm SE_{sps}}$ algorithms, while residual modulator pattern remains in the projections after ${\rm SE_d}$ due to incorrect scatter correction.
The estimated scatter profiles of the cone-beam scan with modulator obtained by the ${\rm {SE_{d}}}$, ${\rm {SE_{rsl}}}$ and ${\rm {SE_{sps}}}$ methods are plotted in Fig. \ref{fig_Catphan_Scatter_Correction_Scatter} (a), where the reference scatter was computed by subtracting the transmission image of the fan-beam scan with modulator from that of the cone-beam scan with modulator. The scatter profiles of the ${\rm {SE_{rsl}}}$ and ${\rm {SE_{sps}}}$ methods are basically coincident, which are closer to the reference ones than that of the ${\rm {SE_{d}}}$ method. In addition, the SPR profiles in 
Fig. \ref{fig_Catphan_Scatter_Correction_Scatter} (b) also indicate that the scatter estimation performance of the ${\rm {SE_{sps}}}$ method is similar to
that of the ${\rm {SE_{rsl}}}$ method. 

\begin{figure}[htb]
	\centering
	\includegraphics[width=15cm]{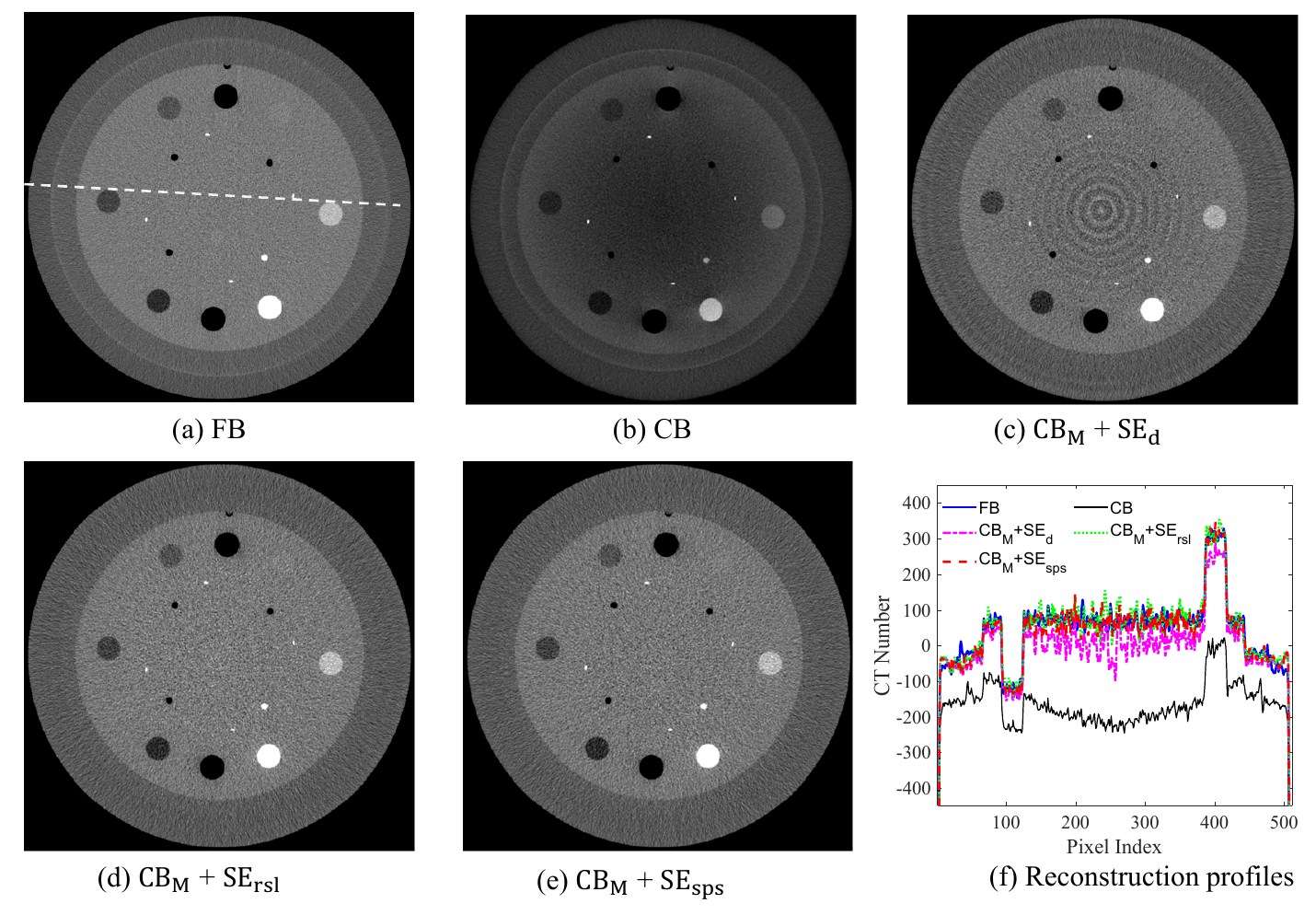}
  \caption[]{Reconstructed CT images of Catphan phantom.
  Reconstruction size: $512\times512$ pixels with $0.4\times0.4$ mm$^2$ pixel size. Display window: [-350, 550] HU.
  (a) CT image from a fan-beam (FB) scan; (b) CT imaged from a cone-beam (CB) scan; (c) CT image from a cone-beam scan with modulator (${\rm CB_M}$) with ${\rm SE_d}$; (d) CT image from a ${\rm CB_M}$ scan with ${\rm SE_{rsl}}$; (e) CT image from a ${\rm CB_M}$ scan with ${\rm SE_{sps}}$; and (f) reconstruction profiles of (a) $\sim$ (e) along the line indicated in (a). 
		\label{fig_Catphan_Scatter_Correction_Recon}  }
\end{figure}

\begin{figure}
	\centering
  \includegraphics[width=15cm]{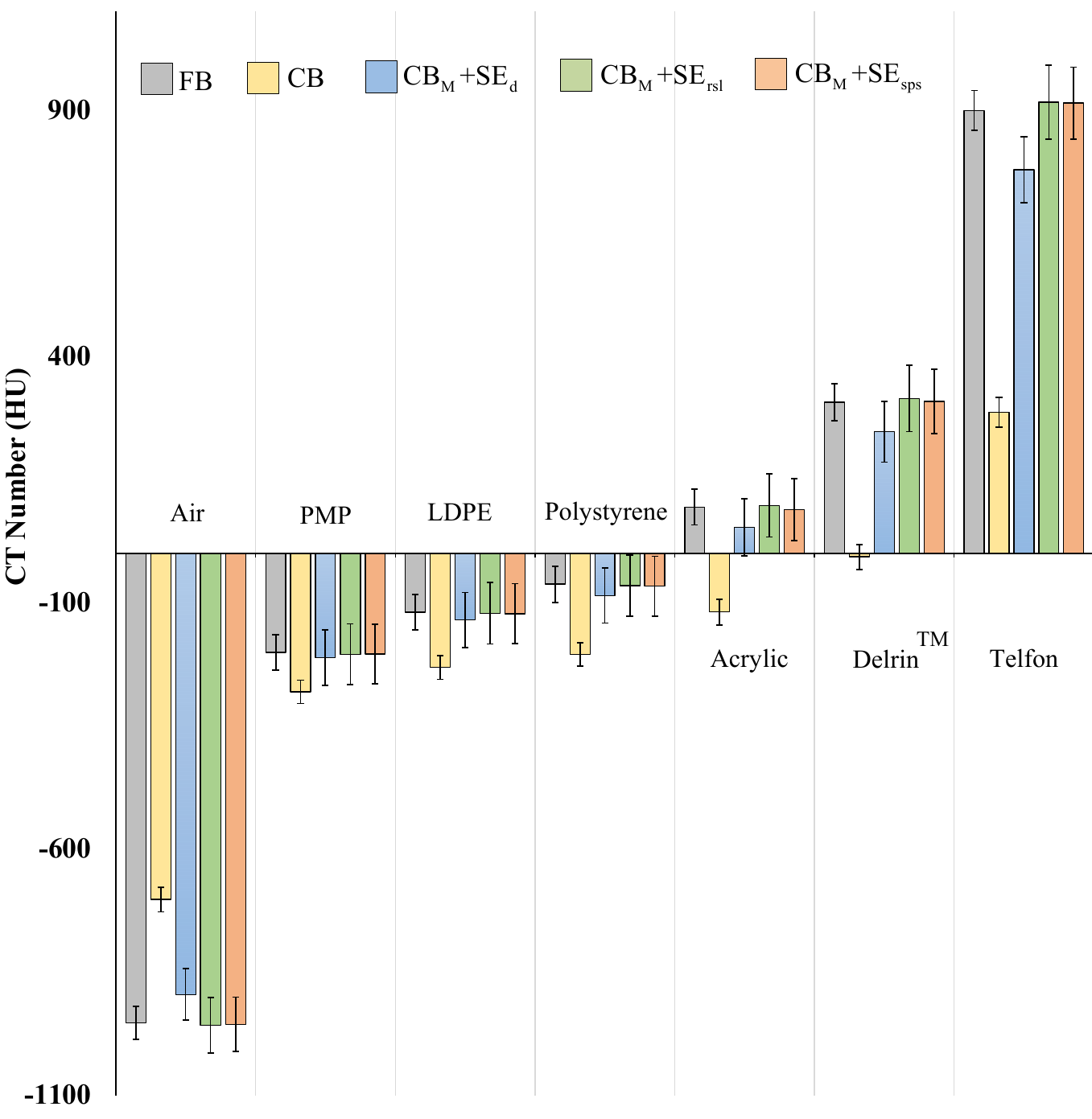}
  \caption[]{Mean and standard deviation of CT number of selected ROIs of Fig.~\ref{fig_Catphan_Scatter_Correction_Recon} (a)$\sim$(e).
		\label{fig_Catphan_Scatter_Correction_ROI}  }
\end{figure}

Reconstructed CT images of the Catphan phantom (no ring correction was applied) are displayed in Fig. \ref{fig_Catphan_Scatter_Correction_Recon}.
Without scatter correction, reconstructed CT numbers from the cone-beam scan is seen to be lower, as expected. The ${\rm {SE_{rsl}}}$ and ${\rm {SE_{sps}}}$ methods can significantly improve the CT number accuracy of the CT image from the cone-beam scan with modulator. The average value and standard deviation of CT number of selected ROIs of Fig.~\ref{fig_Catphan_Scatter_Correction_Recon} (a)$\sim$(e) are demonstrated in Fig.~\ref{fig_Catphan_Scatter_Correction_ROI}. The $E_{{\rm RMSE}}$ of average CT number of selected ROIs was significantly reduced
from 297.9 to 6.5 HU by ${\rm {SE_{sps}}}$, while to 7.7 HU by ${\rm {SE_{rsl}}}$ and 
to 58.5 HU by ${\rm {SE_{d}}}$.

Compared with the ${\rm {SE_{d}}}$ method, it shows that the change of scatter distribution after spectral correction is correctly predicted by our proposed scatter property and should be taken into account in relevant applications such as in estimating scatter from the SMFFS as the ${\rm {SE_{sps}}}$ method does. 
It is worth noting that the ${\rm {SE_{rsl}}}$ and ${\rm {SE_{sps}}}$ method are close mathematically and have similar scatter estimation performance, as demonstrated in the above experiment results. However, from practical application and algorithm implementation point of view, they are quite different. The ${\rm {SE_{rsl}}}$ method is derived by assuming a ``dual-energy'' SMFFS data acquisition. It requires an additional linearization using the residual spectrum but can easily be extended to triple-energy or more. So it is compatible with material decomposition if doable.
On the contrary, the ${\rm {SE_{sps}}}$ method is derived based on a scatter property we proposed here. It needs no additional linearization beyond the conventional beam hardening correction function.

\section{Discussions} \label{sec:Discussions}

X-ray scatter and beam hardening of the spectrum are two major physics challenges in computed tomography. In this work, we aim to understand and characterize the behavior of scatter in CT imaging scenarios where the scatter correction is coupled with the spectral correction. The proposed scatter property says that when applying the spectral correction before scatter is removed, the impact of SPR on the CT projections will be scaled by the first derivative of the mapping function; while the scatter distribution in the transmission domain will be scaled by the product of the first derivative of the mapping function and a natural exponential of the projection difference before and after the mapping.

It is worth noting that in order to characterize the new property in a simplified form, a Taylor expansion and some reasonable approximations were taken, meaning that after spectral corrections both the SPRC-II method and the new scatter distribution  given by this scatter property are approximated. However, our SPR compensation study and evaluations on the scatter correction performance for the SMFFS suggest that such approximations and the proposed scatter property are highly valid, even in situations where SPR values were as high as 300\% as seen with the anthropomorphic chest phantom in Fig.~\ref{fig_Chestphan_SPR_Projection_Recon_120kVp}. 

\begin{figure}[htb]
	\centering
	\includegraphics[width=15cm]{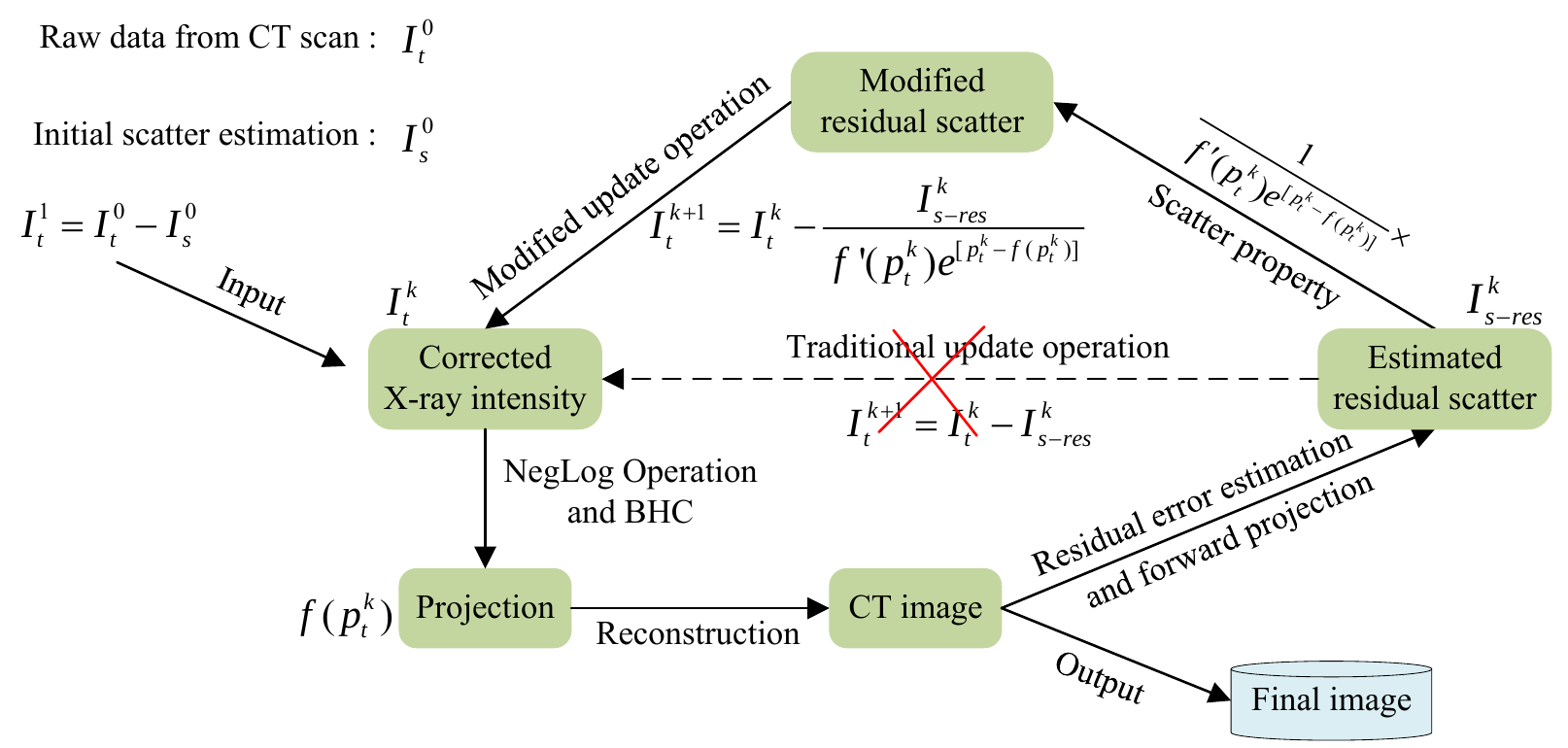}
  \caption[]{A diagram of an inner-loop correction for scatter and beam hardening. 
		\label{fig_Loop_correction}  }
\end{figure}

Our proposed scatter property may also be of help in an iterative approach where scatter and beam hardening are modeled and corrected within a unified framework \citep{nuyts2013modelling}. 
As we know, an initial scatter estimation is usually not accurate enough for a 
complicated object in applications where high image quality is crucial and the 
residual scatter cannot be ignored.
The iterative scheme between projection domain and image domain 
can be leveraged to correct for the residual scatter.
Specifically, as shown in Fig.~\ref{fig_Loop_correction}, in the $k-$th iteration ($k>=1$), the projection data $f(p_{t}^{k})$ is acquired from the X-ray intensity $I_{t}^{k}$
after negative logarithm operation and beam hardening correction, and then the CT image is reconstructed as usual. 
In the image domain, prior knowledge and a regularizer can be applied to 
estimate the impact of residual error from scatter, an idea similar to Ref. \citep{niu2012quantitative}.
Finally, when going back to the transmission domain after forward projection,
the X-ray intensity is updated according to the estimated residual scatter.
In a traditional update operation with no consideration of the impact of beam 
hardening correction to the residual scatter distribution, the estimated residual scatter $I_{s-res}^{k}$ is directly subtracted from X-ray intensity $I_{t}^{k}$.
However, according to our proposed scatter property in this paper, the beam hardening correction will affect the residual scatter distribution. As a result, before the subtraction, $I_{s-res}^{k}$ should be adjusted by a scaling factor of 
$\frac{1}{f^{'}(p_{t}^{k})e^{[p_{t}^{k}-f(p_{t}^{k})]}}$
for each detector pixel based on our analysis.
With more appropriate handling of residual scatter in the utilization of our proposed scatter property, the convergence of such an iterative approach may be improved.
As a matter of fact, whenever a nonlinear operation such as beam hardening correction 
is applied, its impact will propagate to the residual error (e.g., uncorrected scatter) and should be taken into account when a correction for the residual error is done iteratively. This idea will be evaluated in future work.

\section{Conclusions} \label{sec:Conclusions}
In summary, scatter's behavior in X-ray CT spectral correction is characterized, whose correctness and potential were both demonstrated in simulation study as well as real applications for Catphan phantom and the chest phantom on a tabletop CBCT system. In the future, further evaluations and applications of this new scatter property will be conducted, including extending it to other object materials beyond water and water-equivalent spectral correction, as well as deriving forms for situations where two- and multi-material objects with two- and multi-material spectral corrections (e.g., material decomposition) are of concern. 

\section*{ACKNOWLEDGMENTS}
This project was supported in part by a grant from the National Natural Science Foundation of China (No. 12075130 and No. U20A20169) and the New Faculty Startup Fund of Tsinghua University (No. 53331100120).

\section*{APPENDIX A \\
Scatter Estimation method for SMFFS based on Residual Spectral Linearization}

Using the subtraction between Eq.~(\ref{Eq_SMFFS_math_1}) and Eq.~(\ref{Eq_SMFFS_math_2}), the scatter can be eliminated,

\begin{equation}
  I_{t_1}-I_{t_2}=\int S_{res}(E)e^{-u_{o}(E)L}{\rm d} E \label{Eq_SMFFS_res},
  \tag{A1}
\end{equation} 

here, $S_{res}(E)=I_{m_1}S_1(E)-I_{m_2}S_2(E)$, which is the residual spectrum 
between the modulated spectra from flying focal spot positions 1 and 2.
Based on the single-material assumption, the path-length $L$ in Eq.~(\ref{Eq_SMFFS_res}) can be computed by the conventional linearization mapping based on the residual spectrum $S_{res}(E)$. Once path-length $L$ is obtained, one can easily estimate the scatter using Eqs.~(\ref{Eq_SMFFS_math_1}) or (\ref{Eq_SMFFS_math_2}),

\begin{align} 
  I_{s} &=I_{t_1}-I_{m_1}\int S_{1}(E)e^{-u_{o}(E)L}{\rm d} E \nonumber \\ 
        &= I_{t_2}-I_{m_2}\int S_{2}(E)e^{-u_{o}(E)L}{\rm d} E.
        \label{Eq_SMFFS_scatter_rsl}
        \tag{A2}
\end{align}

The scatter estimation method defined in Eqs.~(\ref{Eq_SMFFS_scatter_rsl}) is denoted as ${\rm SE_{rsl}}$.


\end{document}